# Holography inspired self-controlled reconfigurable intelligent surface

Jieao Zhu[1,3], Ze Gu[2,3], Qian Ma[2], Linglong Dai[1]✉, and Tie Jun Cui[2]✉

[1] Department of Electronic Engineering, Tsinghua University, 100084 Beijing, China, and also with the Beijing National Research Center for Information Science and Technology (BNRist), 100084 Beijing, China.
[2] State Key Laboratory of Millimeter Waves, Southeast University, 210096 Nanjing, China.
[3] These authors contributed equally: Jieao Zhu, Ze Gu.
Email: zja21@mails.tsinghua.edu.cn, guze@seu.edu.cn, maqian@seu.edu.cn, daill@tsinghua.edu.cn, tjcui@seu.edu.cn

## Abstract

Among various promising candidate technologies for the sixth-generation (6G) wireless communications, recent advances in microwave metasurfaces have sparked a new research area of reconfigurable intelligent surfaces (RISs). By controllably reprogramming the wireless propagation channel, RISs are envisioned to achieve low-cost wireless capacity boosting, coverage extension, and enhanced energy efficiency. To reprogram the channel, each meta-atom on RIS needs an external control signal, which is usually generated by base station (BS). However, BS-controlled RISs require complicated control cables, which hamper their massive deployments. Here, we eliminate the need for BS control by proposing a self-controlled RIS (SC-RIS), which is inspired by the optical holography principle. Different from the existing BS-controlled RISs, each meta-atom of SC-RIS is integrated with an additional power detector for holographic recording. By applying the classical Fourier-transform processing to the measured hologram, SC-RIS is capable of retrieving the user's channel state information required for beamforming, thus enabling autonomous RIS beamforming without control cables. Owing to this WiFi-like plug-and-play capability without the BS control, SC-RISs are expected to enable easy and massive deployments in the future 6G systems.

# Introduction

Future sixth-generation (6G) communication systems are envisioned to achieve high data rates and wide wireless coverage for emerging applications, such as augmented reality[1], digital twins[2], and autonomous driving[3]. While most advances in the next-generation communication technologies are focused on transceiver designs, another promising approach is to artificially reconfigure the wireless channel for favorable propagation conditions[4-10]. Thanks to recent progress in metamaterials and metasurfaces[5,6,9,11-14], reconfigurable intelligent surfaces (RISs) have been proposed to directly manipulate the channel for enhanced wireless transmissions. Specifically, an RIS is a planar array typically composed of multiple low-cost nearly-passive tunable meta-atoms, which can controllably alter the amplitude, phase, frequency, polarization, and other electromagnetic characteristics of the incident signals. By correctly configuring these meta-atoms, RISs can controllably alter the wireless channel between the base station (BS) and the user equipment (UE), thus overcoming shadowing and deep fading problems commonly encountered in complicated propagation environments, e.g., dense urban areas[10]. Therefore, RISs can extend the coverage[15], save hardware cost[16], and reduce power dissipation[17] of the future 6G wireless systems.

To make RISs correctly function, it is a prerequisite to generate control signals. In real-world wireless applications, the control signals should be adapted to the fast-varying propagation environment[10]. In response to the challenging task of adaptive control signaling, RISs are usually controlled by BS, i.e., this control signal has to be transmitted from BS to RIS in either wireless or wired way. For wireless controls, complicated radio frequency (RF) receivers and extra baseband processors are required at each RIS[18,19], which will eventually neutralize the low-cost benefit of RISs. Thus, wired control is a more common practice in engineering. For the wired control, although additional RF receivers and processors are exempted, it still requires at least one additional control cable per deployed RIS, which limits the remote installation and massive deployments of RISs. To remove the need for control cables in the BS-controlled RISs, some prior works introduced external control methods to metasurfaces, including thermal[12], optical[20], gyroscopic[21], and visual[6,22] control methods. However, none of these methods are suitable for wireless applications, since wireless systems usually do not contain thermal or visual signals. Therefore, to enable massive deployment of RISs, an important problem arises: How to realize cableless RIS control without introducing external control mechanisms?

In this article, we propose a holography-inspired self-controlled RIS (SC-RIS) that no longer needs a control line by leveraging the holographic principle. To compensate for the loss of the control line, the control signal should be self-generated by RIS, which requires the RIS to collect the channel state information (CSI) by itself. To obtain the CSI, each meta-atom of the proposed SC-RIS is additionally equipped with a power detector[23]. Based on the holographic recording principle[24], the entire EM wavefront information can be simultaneously recorded in the hologram captured by the power detectors, thus realizing the CSI acquisition even without an RF receiver. Specifically, in the application scenario of RISs, CSI is mainly determined by the user's angular location. This angular location can be computed by applying a two-dimensional fast Fourier transform (2D-FFT) to the measured hologram. After obtaining the user's angle information, beamforming can be performed by RIS to steer the beams towards the user's angles, thus achieving the self-controlled functionality of RIS. Field test experiments demonstrate that, even without the control line, the accuracy to estimate the angular location using the proposed SC-RIS approaches the theoretical array resolution limit, and the presence of SC-RIS can automatically enhance the average received power by 16.4 dB in the practical wireless communication scenarios. Owing to the self-controlled capability, the proposed SC-RIS is envisioned to work in a plug-and-play manner like domestic Wi-Fi devices. Furthermore, without the controlling cables, SC-RISs can be more easily and massively deployed.

**Principle**

Figure 1 demonstrates the overall schematic of the proposed holography inspired SC-RIS, which is independent of the control by BS, as shown in Fig. 1(a-b). Even without the control cable, SC-RIS is capable of automatically enhancing the communication links. To achieve this self-controllability, it is necessary to sense the electromagnetic (EM) environment in an efficient and low-cost way. Inspired by the optical holographic interference experiment shown in Fig. 1(c), we intentionally create an EM hologram at the SC-RIS by simultaneous microwave illuminations from both BS and user. As shown in Fig. 1(d), the proposed SC-RIS, operating at $f_c = 3.5$ GHz, consists of an array of dual-functional meta-atoms. Like the traditional RIS, each meta-atom works as a space-fed tunable reflective phase shifter. Additionally, the meta-atoms of SC-RIS are equipped with power detectors to record the EM hologram. After the power detectors collect the holographic measurements, these data are transferred to a microcontroller unit (MCU). Since the hologram contains the user's location information, a holographic localization algorithm is executed by MCU to estimate the user angular location. Finally, with the estimated user location, a coding sequence is generated to control the phase shifts of each meta-atom to fulfill collaborative beamforming to arbitrary user, achieving the SC-RIS's closed-loop self-controllability.

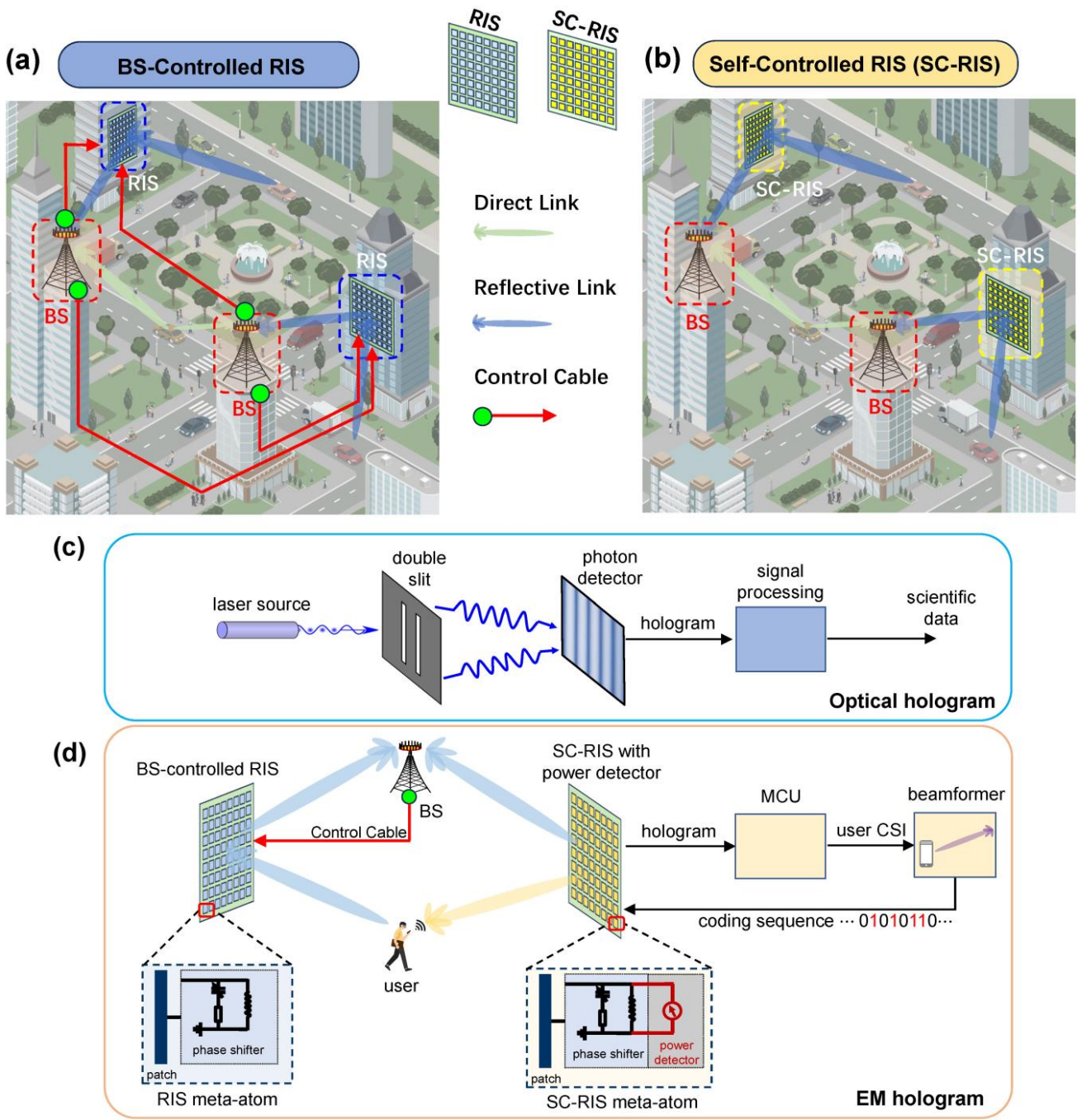

**Fig. 1 Overall schematic of holography inspired self-controlled reconfigurable intelligent surface (SC-RIS).** (a) RISs are usually deployed on urban building facets to dynamically overcome shadowing by providing additional reflective links. Traditional RISs are wire-controlled by the base station (BS), thus incurring a complicated control cable cost. (b) The proposed SC-RIS is capable of "cutting" the control cable and acting as a self-controlled smart

reflective beamformer. (c) Workflow of the optical holography experiment. (d) Inspired by the optical holography, SC-RIS is capable of measuring the incident EM intensity (hologram) via an additional power detector array, and self-configures their phase shifts for beamforming according to the processed hologram.

**Meta-unit design.** To achieve the holographic measuring, it is required to record the entire intensity information on the whole SC-RIS. Thus, the holographic sensing apparatus should be both widely spread across the surface and closely integrated into each meta-atom. To satisfy these two technical needs, we design a hardware-structure multiplexing method in the meta-unit design that successfully solves the problems by integrating the sensing circuit into the meta-unit[23,25], as demonstrated in Fig. 2. Through the integration, the meta-unit constituting the SC-RIS is entitled with a sub-wavelength sensing capability, reaching the same scale as its EM manipulation capability. In Fig. 2(a), we show a simplified diagram of the designed meta-unit with principal illumination. Most of the energy that impinges the meta-unit is reflected with a 1-bit phase modulation empowered by the active resonant structure controlled by two PIN diodes, while a certain proportion of energy is coupled into the detector chips integrated into the meta-unit for sensing the illumination intensity. A diode-control and voltage-read-out circuit is embedded inside the unit, whose routings are carefully designed to avoid the coupling between these two parts of different functionalities, as shown in Fig. 2(b).

Figures 2(c) and 2(d) depict the frequency response of the unit's reflective parameter and detector voltage output. Around the central frequency, a 220 MHz bandwidth is realized for a 180-degree phase difference between the two states within a 20-degree range. The sensitivity of the meta-unit reaches its maximum at the frequency of approximately 3.1 GHz, as shown in the subfigure of Fig. 2(d), which depicts the voltage frequency relationship under a certain injection power level. The minor deviations between the valid frequency bands may arise from detuning inside the matching network. Nevertheless, the discrepancy only leads to a small drop in the detector output voltages, which are still able to be collected by analog-to-digital conversion (ADC) modules. Meanwhile, from the curves in Fig. 2(d), we can tell that the power sensing performance is also limited by the dynamic range of the detectors, while a low-power energy injection could not be identified. In real application scenarios, microwave amplifiers or automatic gain control (AGC) modules will be added, which can amplify the coupled energy while retain the relative intensity magnitudes among different sensing nodes.

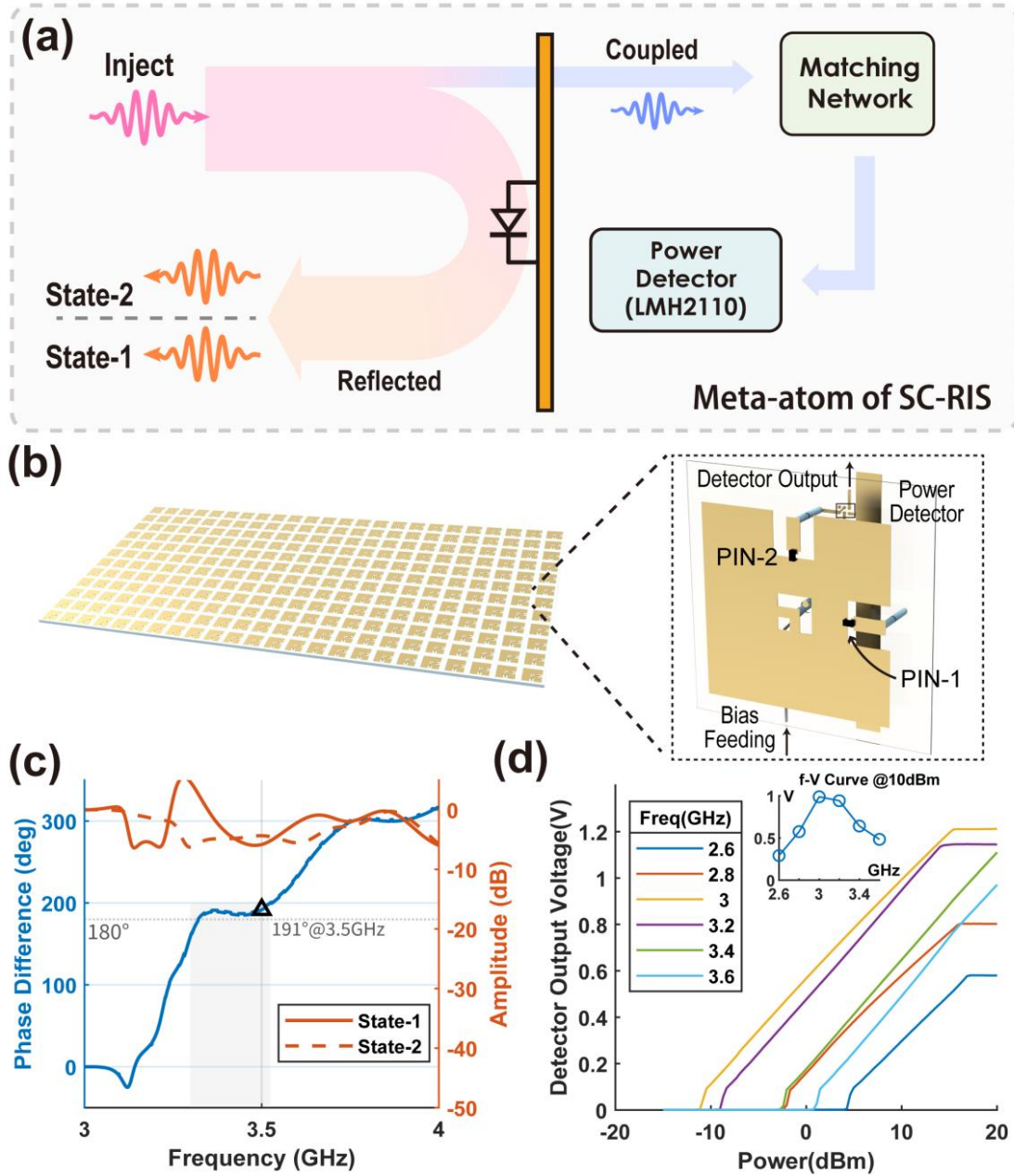

**Fig. 2 Meta-unit with dual electromagnetic sensing-manipulation functionality.** (a) Mechanism of the meta-unit. Similar to a traditional reflective meta-unit, most injected electromagnetic energy is reflected through the unit that has a 1-bit phase modulation ability, marked by State-1 and State-2 in the figure. A fixed amount of injected energy is coupled into the integrated detector inside the unit, passing through a matching network mid-way for impedance matching. Compared with other metasurface systems with a limited sensing ability, the adopted SC-RIS structure integrates the sensing capability to a unit level, increasing the sensing resolution and making holographic self-control possible. (b) A demonstration of the RIS and the magnified model of the unit model. (c) The measured reflective parameters under two states. Around the centering frequency, a 220 MHz bandwidth is realized for a 180-degree phase difference between two states under a 20-degree deviation. (d) The measured detector output voltages of the meta-unit under the EM wave illumination of different intensities and frequencies. In the subfigure, the f-V (frequency vs. voltage) is given for a certain power level, validating the resonant position of the sensing circuit design.

**Algorithm.** To achieve the self-controlled autonomy, the location information of the user is required at SC-RIS. Thus, we design the holographic localization algorithm, which is responsible for extracting the location information of each user from the measured hologram. To create the hologram containing the user location, we allow two coherent EM sources (BS and UE) to simultaneously illuminate the SC-RIS, as is shown in Fig. 3(a). Note that this method strictly follows the holographic imaging principle[24]. On one hand, the EM signal from BS plays the role of reference wave, which is denoted as $\beta$. Since BS is usually installed at a known location in the communication environment, $\beta$ is assumed known to the MCU processor. On the other hand, the EM signal from the user corresponds to the object wave in holographic imaging, which is denoted as $\alpha$. The hologram intensity $I$ is thus created by the EM interference between $\alpha$ and $\beta$ waves, which is represented as

$$I_{mn} = |\alpha_{mn} + \beta_{mn}|^2, \qquad (1)$$

where indices $m$ and $n$ indicate the row and column of the hologram, respectively.

To extract the user CSI from the hologram, it is necessary to study its structural characteristics. Assuming that BS and UE are located in the far-field region of SC-RIS, we analyze the spectral properties of the hologram in the wavenumber domain (k-domain), which can be efficiently computed by FFT (see Supplementary Note 1). Fig. 3(b) illustrates the three-peak spectral characteristics of a simplified one-dimensional hologram. In this spectral representation, the BS signal $\beta$ (red) interferes with the UE signal $\alpha$ (blue), creating a hologram with three peaks in the transformed k-space (pink). Two symmetric peaks are originated from the differential frequencies between the BS and UE signals, and the remaining zero-frequency peak is explained by the total signal energy of the hologram. Since the symmetric peaks are determined by the unknown UE location and the known BS location, the detection of symmetric peaks will help retrieve the UE location.

By exploiting the three-peak spectral structure of the EM hologram, we design a holographic localization algorithm, as elaborated in Fig. 3(c). The algorithm relies on a 2D-FFT analysis on the hologram. Specifically, by expanding Eq. (1), the two squared terms $|\alpha_{mn}|^2 + |\beta_{mn}|^2$ correspond to the zero-frequency part of the holographic spectrum, while the cross terms $\alpha_{mn}^*\beta_{mn} + \alpha_{mn}\beta_{mn}^*$ encodes the interference between the target UE and the reference BS, contributing to the two differential frequency peaks. By applying a spatial FFT, these interfering terms are converted into the differential frequencies in the k-space, which is expressed as

$$\hat{I}_{k\ell} = \mathcal{FFT}[I_{mn}]_{k\ell} = A\delta_{k\ell}(0) + \left(C\delta_{k\ell}(\Delta_z,\Delta_x) + C^*\delta_{k\ell}(-\Delta_z,-\Delta_x)\right) \qquad (2)$$

where $\Delta_z$, $\Delta_x$ are spatial frequency differences of $\alpha$ and $\beta$ in the $z$, $x$-directions, respectively. The zero-frequency term in (2) represents the average intensity $A$ of the hologram, and the second term represents the spatial varying structure of the hologram. Thus, by performing a 2D-FFT on the hologram $I_{mn}$, we can retrieve the spatial frequency differences $\Delta_z, \Delta_x$ from the spectral peaks in $\hat{I}_{k\ell}$. Since the spatial frequencies are uniquely determined by the user's incident angles, we can finally compute the angular location $(\theta_{\text{UE}}, \varphi_{\text{UE}})$ of the unknown target from the observed hologram (see Methods for algorithm details). However, the spectral peaks in $\hat{I}_{k\ell}$ appear in pairs, indicating that there are two candidate solutions to the user localization problem. This is caused by the indistinguishability between the two interfering cross terms $\alpha_{mn}^*\beta_{mn} + \alpha_{mn}\beta_{mn}^*$. This ambiguity is known as the *twin image problem* in optical holography[24]. Fortunately, the twin image can be removed by the regulated compressed sensing-based algorithms[26] or by applying practical physical constraints to the reference wave[27] (see Supplementary Note 2). Thus, in this work, we generally assume that one out of the two peaks representing the true BS-UE relative direction can be correctly identified.

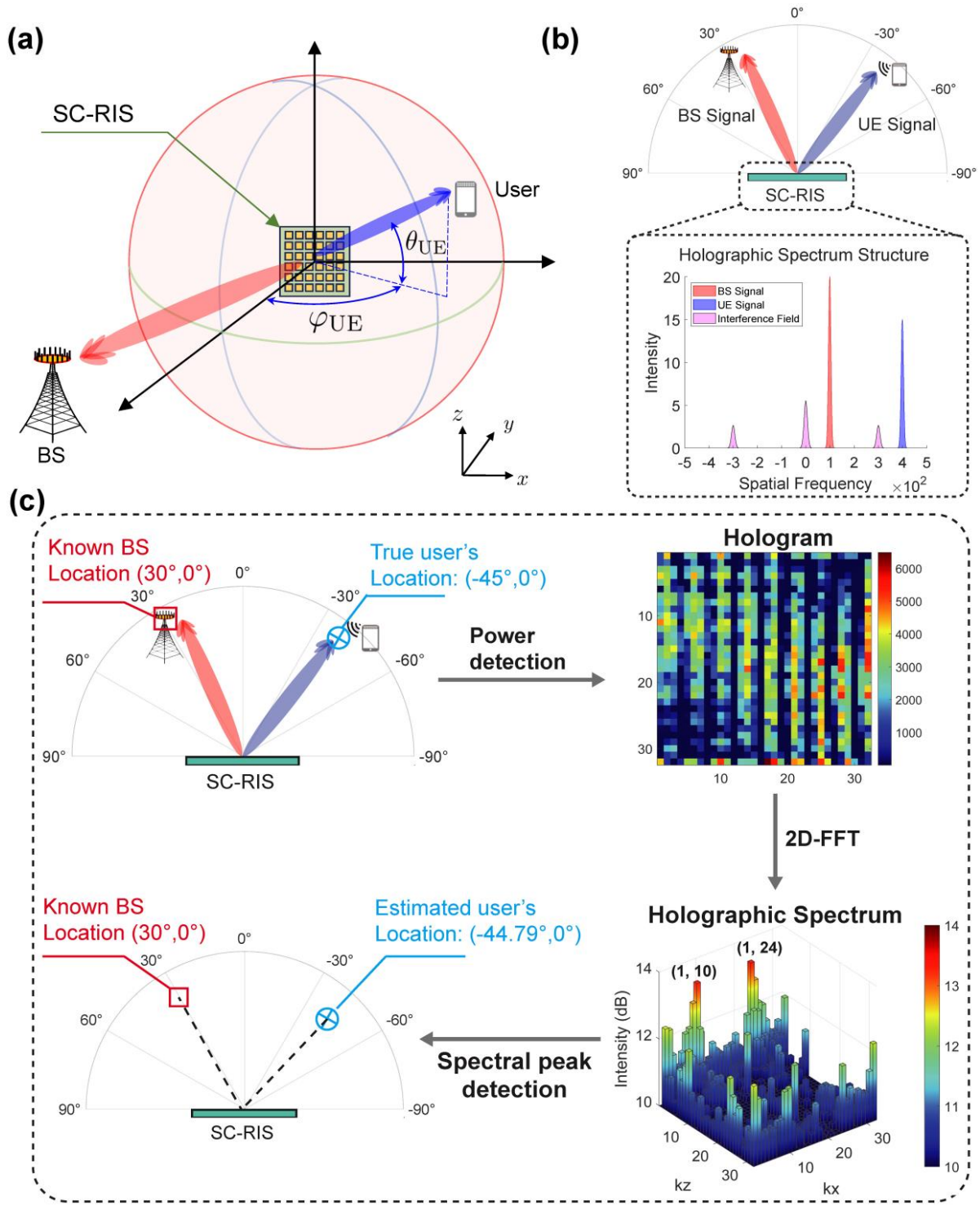

**Fig. 3 Holographic localization algorithm.** (a) Coordinate system for analysis of SC-RIS. The SC-RIS is placed in the xOz plane, centered at the origin, where the unknown location of the UE is described in spherical coordinates. (b) Spectral characteristics of the hologram, viewed along the x-axis. In the wavenumber domain (k-space), the hologram contains three spectral peaks: the two symmetric differential frequencies that arise from mutual-interfering terms, and the zero-frequency peak that comes from the self-interfering term. (c) Schematic diagram of obtaining user's CSI by the holographic localization algorithm. To exploit the three-peak structure of the holographic spectrum, a two-dimensional fast Fourier transform (2D-FFT) is applied to the hologram. The angular location of the user can be determined from the differential frequency spectral peaks.

The simulated accuracy of the 2D-FFT algorithm is provided in Supplementary Note 3, which has been compared with high-complexity super-resolution algorithms such as maximum-likelihood (ML) estimators. Furthermore, the robustness of the algorithm against the BS-UE power unbalance is studied and presented in Supplementary Note 4.

## Experimental results

Fig. 4 demonstrates the experimental setup of the holographic localization and corresponding results. A 32×32-unit SC-RIS is firstly fabricated with an overall size of 0.64×0.64m$^2$ and placed in the $xz$ plane for the EM sensing and channel augment, as shown in Fig. 4(a). In the meantime, two terminals, i.e., UE and BS, are characterized by two antennas that target the SC-RIS as independent radiation sources. To be specific, the UE terminal represents the moving target that to be localized and the BS terminal represents the stationary reference point whose location is prior information to the algorithm. The communication link between the UE and BS is also established, where the UE terminal becomes a receiver that collects the signal generated by BS. In the built-up system, an RF SP2T switch is connected to the UE antenna that guides the energy direction from the signal generator (in localization mode) or into the spectrum analyzer (in communication mode). Two modes alternate between each other temporally, constructing a continuous detecting-targeting loop for moving terminals.

A three-level hierarchy for the proposed system is plotted in Fig. 4(b) for a plain demonstration. At the bottom layer, i.e. application layer, algorithms of the proposed non-coherent localization as well the coding generation for SC-RIS are implemented on the processer unit. A set of peripherals inside the processor manages the data flow from or towards the peripheral modules, including an ADC array, a diode driver circuit, and a GUI display in the control layer. In the experiments, we firstly use single-tone signals as the excitations for two terminals. Without loss of generality, the single-tone excitation represents the simplest and most efficient way for interference generation and could be easily analyzed by the spectrum analyzer quantitatively in the simulated communication for channel estimation. In the transmission layer, the antennas are positioned in a fashion that no direct line of sight (LoS) path exists in an analog of obstacle scenarios. The SC-RIS acts as the spatial modulator for channel augment inside the communication link.

To facilitate the result presentation, the middle variables inside the algorithms are collected and demonstrated in Fig. 4 together with the final localization results. Three samples are chosen, where the BS and UE are located at three different locations respectively. As a demonstration for non-coherent holographic detection, the proposed localization algorithm only relies on the intensity information inside the monitoring plane, which is plotted in Fig. 4(c). From the figures, we could clearly observe the interference patterns, i.e., the hologram, that marks the periodical amplitude fluctuation along the y-axis. Based on the principle of the holographic reconstruction algorithm, the peaks of the Fourier-transformed hologram represent the relative angular position of the two radiation points, as shown in Fig. 4(d). The peaks in the spectrum exist in pairs, which matches the two possible solutions caused by the twin image problem, which is previously explained in the algorithm description. Finally, the predicted locations of the UE terminal are derived and demonstrated in Fig. 4(e). The actual positions are given for reference, where the localization error is around $4°$ for the three samples.

The experimental results for channel augment are also provided in Fig. 4. Two methods are introduced for coding generation based on the predicted UE location, i.e. near-field coding mechanism and far-field coding mechanism (see Methods for detailed information). In Fig. 4(f), the near-field codings are adopted for their better EM energy converging performance. The expenses of this method include a higher calculation cost in coding generation and a demand for distance information. However, such defects could be overcome with an introduction of high-performance processors and parallel architecture for distance estimation algorithms, which have been studied extensively[28]. The field converging capability is demonstrated through the focusing patterns shown in Fig. 4(g), where the transmission parameter between the UE and BS terminals is improved by at most 20.1 dB.

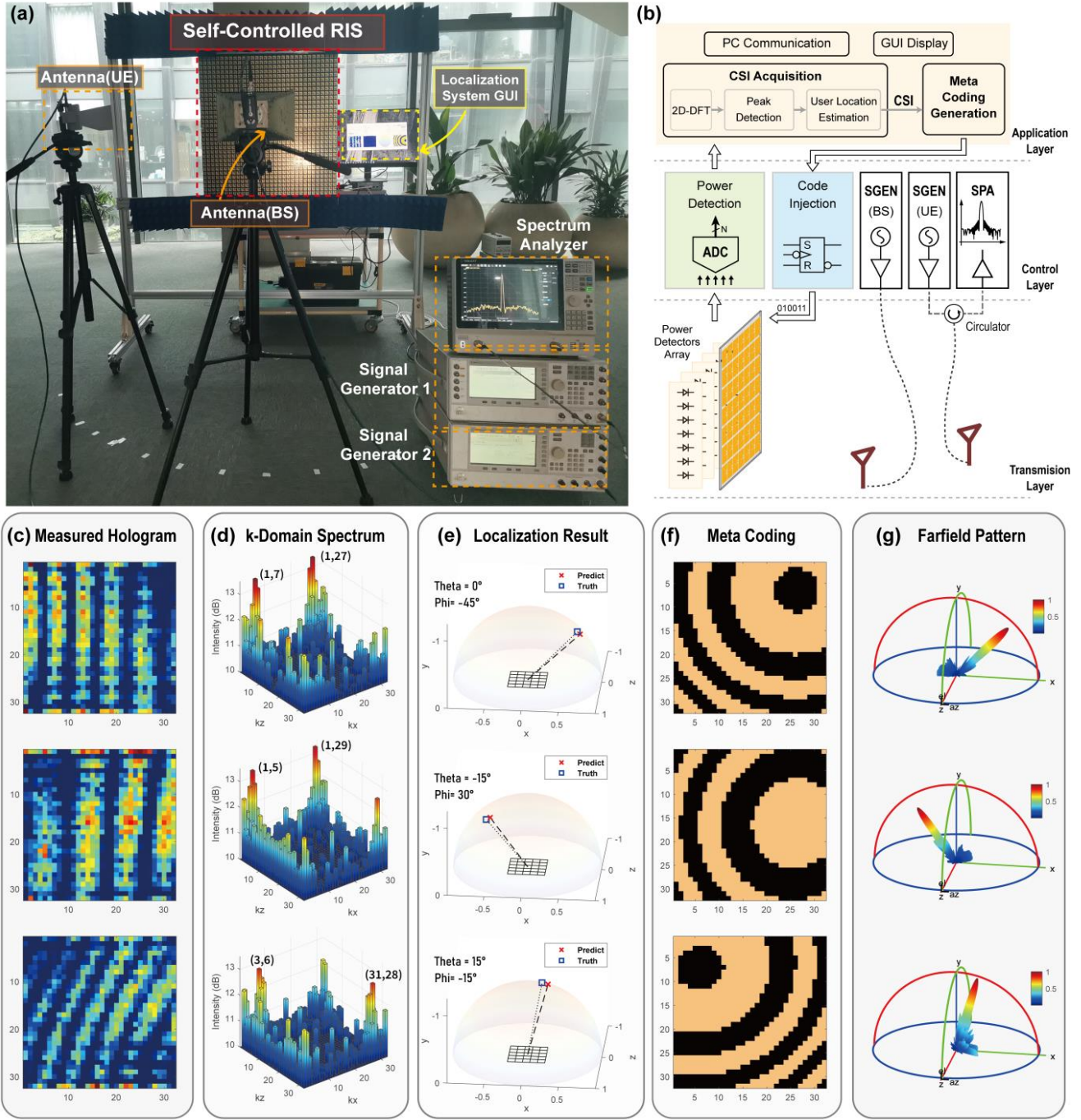

**Fig. 4 Holographic Localization.** (a) Picture of the system. Two radiation terminals are characterized by two antennas, which represent a user that is to be localized (UE) and a fixed-position base station (BS). Each terminal is connected to a signal generator independently. In the localization experiment, single-tone signals of the same frequency are excited from the antennas. The SC-RIS monitors the interference field dynamically via the integrated sensing nodes and updates its coding for channel enhancement according to the *in-situ* localization result. (b) The system diagram for simultaneous localization and channel-augment. (c) Detector array outputs. Three samples are chosen for demonstration. As a demonstration for non-coherent detection, the proposed localization algorithm only relies on the intensity information inside the monitoring plane, which significantly simplifies the system. (d) The k-domain spectrums after 2D-FFT, where the amplitude peaks that represent the interference can be clearly identified. (e) The predicted location of the user. The actual position is given for reference. (f) Optimized coding of SC-RIS for point-to-point communication between UE and BS. (g) The far-field patterns with focusing beams targeted at the UE position after the channel augment.

Following the procedures in Fig. 4(b), the overall system can realize the non-coherent detection and automatic encoding through a self-programming metasurface without any human intervention. A video is presented as a solid demonstration of such autonomous procedures (see Supplementary Information Video 1). Before carrying out the real communication experiments, a comprehensive analysis is firstly presented concerning the system performance.

**Automatic user localization and beamforming.** To quantify the CSI acquisition capability of the proposed SC-RIS, we show the estimation accuracy of the user's angular location $\theta_{UE}, \varphi_{UE}$ by running the proposed holographic localization algorithm on MCU to process the measured data in real time. In the testing scenario, the BS antenna is placed at four different possible locations $(\theta, \varphi)_{BS}$ of $(0°, 0°)$, $(-15°, 0°)$, $(-15°, -30°)$, and $(0°, -30°)$; while the UE antenna is placed at locations $(\theta, \varphi)_{UE}$ of $\theta_{UE} \in \{0°, \pm15°\}$, and $\varphi_{UE} \in \{0°, \pm15°, \pm30°, \pm45°, \pm60°\}$, except for the locations that coincide with BS. For each allowable BS-UE location configuration, the hologram is digitally measured by SC-RIS, and the holographic localization algorithm described in the ***Algorithm*** section is executed to obtain the estimated UE location $(\theta, \varphi)_{est}$.

The localization performance is demonstrated in Fig. 5(a) when the user is fixed at $\varphi_{UE} \in \{15°, 30°, 45°, 60°\}$. Data is gathered by varying the location of the BS antenna and taking multiple observations for each BS-UE configuration. From Fig. 5(a), we observe that the holographic localization algorithm returns angle estimates with error $\leq 9°$ with high probability, coinciding with the simulated results in the algorithm description section. Fig. 5(b) illustrates the estimated user's elevation angle $\theta_{est}$ with respective to SC-RIS, compared with the ideal estimator (dashed line). The shading represents the $\pm 1\sigma$ region. Similarly, the estimation results for the azimuth angle $\varphi$ are shown in Fig. 5(c). Since the manufactured SC-RIS is a square array with $x$ and $z$ symmetry, the estimated results in Fig. 5(b-c) are also symmetric relative to the central point $(0°, 0°)$.

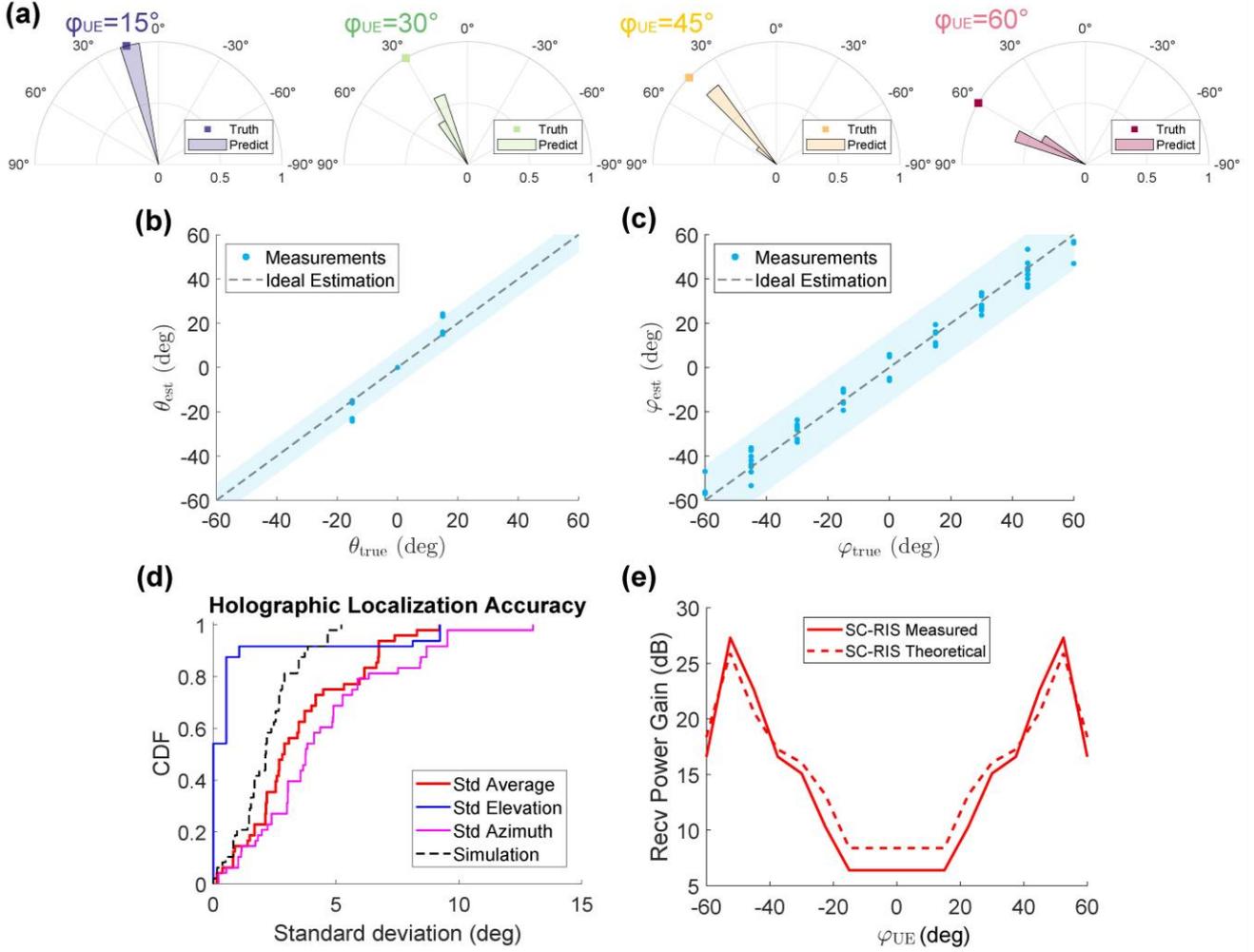

**Fig. 5 System performance of SC-RIS.** (a) Distribution of the SC-RIS localization results with users at different angular locations. (b-c) Measured SC-RIS localization performance, compared with ideal estimators (dashed) which produce estimated angles $\varphi_{est}, \theta_{est}$ that are identical with the true values $\varphi_{true}, \theta_{true}$. (d) Measured localization results compared with simulation. (e) The user's received power gain after activating the SC-RIS. An average improvement of 16.4 dB is observed when SC-RIS is enabled.

To statistically characterize the performance of holographic localization, we further compute the distribution of error for the angular estimators. The standard deviation (STD) is defined as $\text{STD}(\hat{\varphi}) = \sqrt{\frac{1}{N-1}\sum_{i=1}^{N}(\hat{\varphi}_i - \varphi_{\text{true},i})^2}$, where $N$ is the number of experimental data. STD values of 2.61° for the elevation angle and 5.24° for the azimuth angle are observed, compared to the theoretical resolution limit of 3.98° for a $32 \times 32$-element array on each dimension. Since the azimuth angle is estimated after the elevation angle in the 2D-FFT algorithm (See Methods for details), the azimuth deviation is larger than the elevation deviation due to error accumulation. The total average deviation of the proposed estimator is 5.85°, which well approaches the 3.98° theoretical limit. The small error of the 2D-FFT holographic localization algorithm is attributed to the noise of the power detector and the imperfect carrier synchronization. The cumulative distributions of the measured estimation errors are shown in Fig. 5(e), together with that of the simulated data obtained by performing 2D-FFT on the noiseless theoretical interference pattern (dashed). By comparing the red measured curve with the dashed simulated curve, we conclude that the proposed user localization algorithm shows comparable performance against the ideal simulated case, even though the measurements are influenced by the detector noise and imperfect synchronization.

To further verify the automatic beamforming capability of the proposed SC-RIS, we conduct fully automated experiments with the MCU-executed holographic detection and beamforming pipeline. The measured results shown in Fig. 5(f) exhibit an average improvement of 16.4 dB in the user received power when the automatic encoding on SC-RIS is activated. The received power is consistently improved across the azimuth angles of $-60 \sim 60°$, showing a stable reflective gain of the incident continuous microwave without data modulation. The data trend of the measured gain well coincides with the theoretical predictions. Notice that the theoretical predictions are compensated by taking into consideration the insertion losses of the signal generators, antennas, and circulators involved in the field-test experiments.

**Application to real-world communication systems.** To demonstrate the communication-enhancing functionality of SC-RIS, we test the autonomous beamforming function with two USRPs operating at 3.5 GHz, simulating a real-world communication system. Here, the SC-RIS is configured to the auto-beamforming mode, indicating that it is capable of enhancing the BS-user communication link without the control cable. Fig. 6(a) shows the wireless communication experimental settings, in which the data transmission is aided by the fabricated SC-RIS. The SC-RIS is placed in front of the Tx and Rx antennas to provide a controllable reflective link. We test the data transmissions with 1.5-Mbit PNG-format logos of Tsinghua University and Southeast University in Fig. 6(b), under a transmission power of 6 dBm.

To show the autonomous link enhancement capability of SC-RIS, we transmit the same picture data with different configurations of SC-RIS. In the left part of Fig. 6(b), the data are transmitted with all-zero coding, mimicking a regular reflection following the Snell's law. Since real-world Tx and Rx antennas are usually not located along the specular reflection path, the received signal-to-noise ratio is relatively low, resulting in poor quality of the received pictures. To verify the link enhancement of SC-RIS, we firstly configure the Tx and Rx USRPs to transmit a single-tone probe wave, creating an interference pattern on SC-RIS. The interference pattern is captured and then automatically gathered by MCU for holographic localization. After computing the user's location, SC-RIS is auto-configured to steer the reflective signal towards the user. The right part of Fig. 6 shows the received symbols and decoded pictures with the aid of SC-RIS, exhibiting improved picture resolutions.

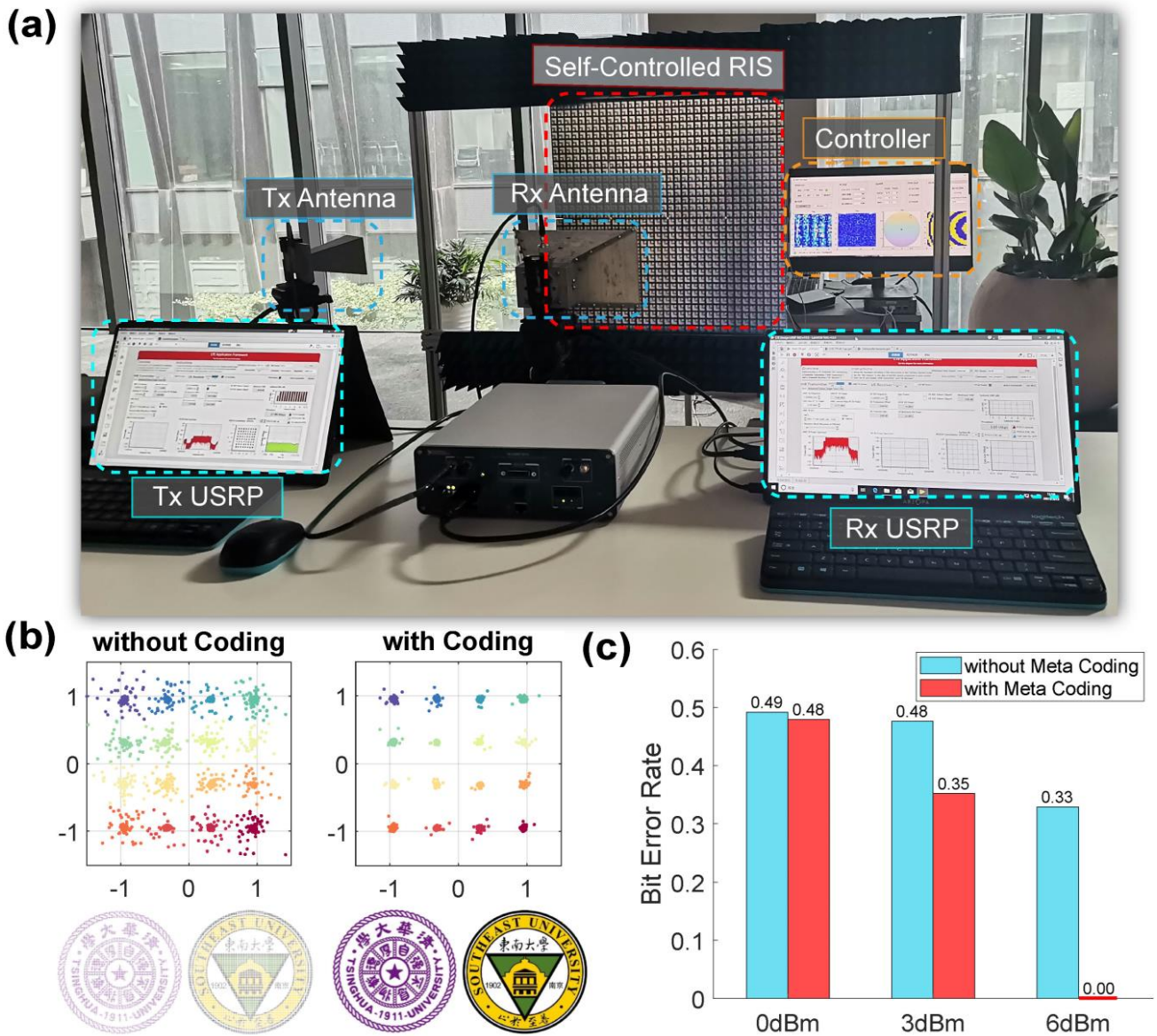

**Fig. 6 Autonomous beamforming of SC-RIS in a wireless communication system.** (a) Photograph of the SC-RIS-aided communication system. (b) Received information-carrying communication symbols, in which the Tsinghua University and Southeast University logos are transmitted. The symbol noise is significantly reduced when the autonomous beamforming function is activated, and the resolution of the reconstructed picture is improved simultaneously. (c) Bit error rates measured at the receiver with the transmit power ranging from 0 dBm to 6 dBm.

To quantify the communication performance improvement brought by SC-RISs, we analyze the bit error rates (BERs) with and without SC-RIS in Fig. 6(c). The error rates are computed within consecutive transmission blocks of 12 Mbit size. Since the error rates of communication systems usually exhibit a waterfall effect (the error rate suddenly drops as the transmitted power increases over some threshold value), we gradually increase the transmitting power from 0 dBm to 6 dBm, with a 3 dB step. When the SC-RIS is not activated, the error rate drops slowly as the transmitted power increases. However, when we repeat the experiment with the SC-RIS activated, the error rate exhibits an obvious waterfall at 3 dBm~6 dBm transition of the transmitted power, as shown in Fig. 6(c). Apart from picture and data transmissions, a video transmission experiment is also conducted for long-term stability tests of the entire system (see Supplementary Information Video 2). The solid experiments on communication systems have demonstrated the autonomous link-enhancing capability of the proposed SC-RIS without the BS control.

## Discussions

**Multi-user localization.** In this article, single-user localization is achieved by the BS-UE interference on SC-RIS. In commercial communication systems, usually dozens of users are served by a single BS simultaneously. A wiser approach is to solve dozens of localization problems at the same time, which we name *multi-user localization*. Inspired by the orthogonal frequency-division multiple access (OFDMA)[29] technology that has been widely applied in the wireless communications, $K$ users can directly transmit the EM signals towards the SC-RIS for simultaneous multi-user localization, as long as these users transmit signals that are separate in the frequency domain.

**Connection with other holography-related technologies.** The simple yet powerful holographic principles enable various applications in interferometric microscopy[30], interference lithography[31], acoustic holography[32], and so on. These specific applications achieve their advantages mainly due to three reasons. Firstly, it is because the holographic recordings contain complete information on the wavefront, enabling efficient subsequent information processing and intelligent decision-making in complicated environments. In the specific case of information metasurfaces[33-37], since microwave signals experience fading, random scattering, and mutual coupling in the complex propagation environment, the wavefront is distorted unpredictably, thus making dynamic adaptation of the microwave information metasurfaces a key challenge. With the help of SC-RISs, it is possible to treat these imperfections altogether as a distortion of the wavefront, which is holographically recorded in a physical manner. Secondly, it is because non-coherent intensity detectors are usually easy to realize, or in some cases it is the only choice, for example, electron holography. Although non-coherent detectors require the known global reference wave, this requirement is easily met in the actual SC-RIS-based systems, since the coherent microwave illumination is common practice and is typically realized by carrier synchronization. The third reason is the high-performance holographic signal processing directly in the microwave domain[33]. Owing to the recent advances in microwave neural networks, microwave signals can be transformed and detected at the speed of light, thus achieving a much higher processing speed of the holography compared to the traditional MCU implementations. In this way, the SC-RIS may respond much quicker to the environmental changes.

**Future work on improving array resolution.** In this article, we have realized the holographic CSI acquisition mainly through the estimation of user's angular location, achieving a measured angular resolution of 5.85° with a $32 \times 32$ power detector array. Although approaching the theoretical resolution limit, this measured resolution is still not satisfactory for real-world wireless applications. Fortunately, this defect can be mitigated by aggregating measured data in the time domain. Generally, with the number of snapshots being $T$, the estimator variance will be reduced by a factor of $\mathcal{O}(1/T)$. Another approach is to enlarge the array size for spatial diversity gain, which scales in the same law with its time-domain counterpart.

## Methods

**Numerical algorithm.** The 2D-FFT is implemented by two consecutive 1D-FFTs, which are applied consecutively to the two indices of the input hologram $I_{mn}$. The overall 2D-FFT-based holographic localization algorithm is summarized in **Algorithm 1**. To overcome the signal aliasing caused by the periodicity of the FFT algorithm, we introduce the *regulate* function in the algorithm to ensure the physical constraint $k_x^2 + k_z^2 \leq k_0^2$ on wavenumbers. After the regulation, the location of the user is calculated by

$$\theta_{\text{UE}} = -\sin^{-1}\frac{\omega_z}{2\pi}, \tag{3}$$

$$\varphi_{\text{UE}} = \sin^{-1}\left(\frac{\omega_x}{2\pi \cos\theta_{\text{UE}}}\right).$$

---

**Algorithm 1:** 2D-FFT Holographic User Localization

**Inputs:**

The measured hologram $I_{mn} \in \mathbb{R}^{N_z \times N_x}$, power detector spacing $d_z, d_x$, operating wavelength $\lambda$, BS angular location $\theta_{\text{BS}}, \varphi_{\text{BS}}$.

**Outputs:**

Estimated user's location $\theta_{\text{UE}}, \varphi_{\text{UE}}$.

1:    $\omega_{x,\text{BS}} \leftarrow 2\pi(d_x/\lambda)\cos\theta_{\text{BS}}\sin\varphi_{\text{BS}}$

2:    $\omega_{z,\text{BS}} \leftarrow -2\pi(d_z/\lambda)\sin\theta_{\text{BS}}$

3:    **Compute** 1D-FFT on index $m$: $\check{I}_{kn} \leftarrow \text{FFT}[I_{mn}]$, $\forall n \in \{1, \cdots, N_x\}$

4:    **Compute** 1D-FFT on index $n$: $\hat{I}_{k\ell} \leftarrow \text{FFT}[\check{I}_{kn}]$

5:    **Find** the spectral peak of the hologram: $(i_z, i_x) \leftarrow \arg\max_{(k,\ell)\neq(0,0)} |\hat{I}_{k\ell}|$

6:    $\omega_z^{(1)} \leftarrow 2\pi(i_z - 1)/N_z$, $\omega_x^{(1)} \leftarrow 2\pi(i_x - 1)/N_x$

7:    $\omega_{z,\text{UE}}(1) \leftarrow \text{regulate}\left(\omega_{z,\text{BS}} + \eta_z^{(N_T+1)}\right)$

8:    $\omega_{x,\text{UE}}(1) \leftarrow \text{regulate}\left(\omega_{x,\text{BS}} + \eta_x^{(N_T+1)}\right)$

9:    $\omega_{z,\text{UE}}(2) \leftarrow \text{regulate}\left(\omega_{z,\text{BS}} - \eta_z^{(N_T+1)}\right)$

10:   $\omega_{x,\text{UE}}(2) \leftarrow \text{regulate}\left(\omega_{x,\text{BS}} - \eta_x^{(N_T+1)}\right)$

11:   **Compute** two candidate user's locations $(\theta_{\text{UE}}, \varphi_{\text{UE}})$ by Eq.(3).

**Subroutine** regulate($x$)

R1:   **return** $x - 2\pi \cdot \text{round}(x/2\pi)$

---

**Communication experiment.** In the communication experiment, we adopt two NI USRP devices for data transmission tests. The Tx and Rx USRPs work at the carrier frequency $f_c = 3.5$ GHz with the communication bandwidth $B = 30$ MHz. During the picture transmissions and video transmissions, the physical layer transmitter adopts a standard LTE protocol with modulation and encoding scheme (MCS)-17 (64QAM, $R_{\text{code}} \approx 0.43$), and the receiver is always configured to the same MCS according to the transmitter. Since the USRP provides UDP socket to the application layer, we use Python to access these data sockets and transmit the data payload. During the constellation gathering, MCS is set to MCS-16 (16QAM, $R_{\text{code}} \approx 0.64$) for clarity of the constellation display.

# References

1   Saad, W., Bennis, M. & Chen, M. A vision of 6G wireless systems: Applications, trends, technologies, and open research problems. *IEEE Network* **34**, 134-142 (2019).

2   Khan, L. U., Saad, W., Niyato, D., Han, Z. & Hong, C. S. Digital-twin-enabled 6G: Vision, architectural trends, and future directions. *IEEE Comm. Mag.* **60**, 74-80 (2022).

3   Yang, B. *et al.* Edge intelligence for autonomous driving in 6G wireless system: Design challenges and solutions. *IEEE Wirel. Comm.* **28**, 40-47 (2021).

4   Wang, Y., Lu, H. & Sun, H. Channel estimation in IRS-enhanced mmWave system with super-resolution network. *IEEE Commun. Lett.* **25**, 2599-2603 (2021).

5   Li, L. *et al.* Electromagnetic reprogrammable coding-metasurface holograms. *Nat. Commun.* **8**, 197 (2017).

6   Li, W. *et al.* Intelligent metasurface system for automatic tracking of moving targets and wireless communications based on computer vision. *Nat. Commun.* **14**, 989 (2023).

7   Li, L. *et al.* Intelligent metasurfaces: Control, communication and computing. *eLight* **2**, 7 (2022).

8   Ye, J., Guo, S. & Alouini, M.-S. Joint reflecting and precoding designs for SER minimization in reconfigurable intelligent surfaces assisted MIMO systems. *IEEE Trans. Wireless Commun.* **19**, 5561-5574 (2020).

9   Zhao, H. *et al.* Metasurface-assisted massive backscatter wireless communication with commodity Wi-Fi signals. *Nat. Commun.* **11**, 3926 (2020).

10  Pei, X. *et al.* RIS-aided wireless communications: Prototyping, adaptive beamforming, and indoor/outdoor field trials. *IEEE Trans. Commun.* **69**, 8627-8640 (2021).

11  Zhang, X. G. *et al.* An optically driven digital metasurface for programming electromagnetic functions. *Nat. Electron.* **3**, 165-171 (2020).

12  Aydin, K. *et al.* Frequency tunable near-infrared metamaterials based on VO2 phase transition. *Opt. Express* **17**, 18330-18339 (2009).

13  Alamzadeh, I., Alexandropoulos, G. C., Shlezinger, N. & Imani, M. F. A reconfigurable intelligent surface with integrated sensing capability. *Sci. Rep.* **11**, 20737 (2021).

14  Gu, Z., Ma, Q., Gao, X., You, J. W. & Cui, T. J. Classification of metal handwritten digits based on microwave diffractive deep neural network. *Adv. Opt. Mater.* **12**, 2301938 (2024).

15  Zeng, S., Zhang, H., Di, B., Han, Z. & Song, L. Reconfigurable intelligent surface (RIS) assisted wireless coverage extension: RIS orientation and location optimization. *IEEE Comm. Lett.* **25**, 269-273 (2021).

16  Jian, M. *et al.* Reconfigurable intelligent surfaces for wireless communications: Overview of hardware designs, channel models, and estimation techniques. *Intelligent and Converged Netw.* **3**, 1-32 (2022).

17  You, L. *et al.* Energy efficiency and spectral efficiency tradeoff in RIS-aided multiuser MIMO uplink transmission. *IEEE Trans. Signal Process.* **69**, 1407-1421 (2021).

18  Alamzadeh, I. & Imani, M. F. Sensing and reconfigurable reflection of electromagnetic waves from a metasurface with sparse sensing elements. *IEEE Access* **10**, 105954-105965 (2022).

19  Basar, E. & Poor, H. V. Present and future of reconfigurable intelligent surface-empowered communications. *IEEE Signal Process Mag.* **38**, 146-152 (2021).

20  Zhang, X. G. *et al.* Optoelectronic metasurface for free-space optical-microwave interactions. *ACS Appl. Mater.*

*Interfaces* **15**, 22744–22751 (2023).
21  Ma, Q. *et al.* Smart metasurface with self-adaptively reprogrammable functions. *Light: Sci. & Appl.* **8** (2019).
22  Jiang, S., Hindy, A. & Alkhateeb, A. Sensing aided reconfigurable intelligent surfaces for 3GPP 5G transparent operation. *IEEE Trans. Commun.* **71**, 6348-6362 (2023).
23  Ma, Q. *et al.* Smart sensing metasurface with self-defined functions in dual polarizations. *Nanophotonics* **9** (2020).
24  Benton, S. A. & Bove Jr, V. M. *Holographic imaging*.   (John Wiley & Sons, 2008).
25  Zhu, J. *et al.* Sensing RISs: Enabling dimension-independent CSI acquisition for beamforming. *IEEE Trans. Inf. Theory* **69**, 3795-3813 (2023).
26  Zhang, W. *et al.* Twin-image-free holography: A compressive sensing approach. *Phys. Rev. Lett.* **121**, 093902 (2018).
27  Latychevskaia, T. & Fink, H.-W. Solution to the twin image problem in holography. *Phys. Rev. Lett.* **98**, 233901 (2007).
28  Zhou, T. *et al.* Short-range wireless localization based on meta-aperture assisted compressed sensing. *IEEE Trans. Microwave Theory Tech.* **65**, 2516-2524 (2017).
29  Yin, H. & Alamouti, S. OFDMA: A broadband wireless access technology. *2006 IEEE Sarnoff Symp.*, 1-4 (2006).
30  Ralston, T. S., Marks, D. L., Scott Carney, P. & Boppart, S. A. Interferometric synthetic aperture microscopy. *Nat. Phys.* **3**, 129-134 (2007).
31  Cheng, L. & R. H., L. Interference lithography: a powerful tool for fabricating periodic structures. *Laser & Photonics Reviews* **4** (2010).
32  Ryuji, H., Giorgos, C., Diego Martinez, P. & Sriram, S. High-speed acoustic holography with arbitrary scattering objects. *Sci. Adv.* **8** (2022).
33  Gao, X. *et al.* Programmable surface plasmonic neural networks for microwave detection and processing. *Nat. Electron.* **6**, 319-328 (2023).

## Acknowledgments

This work was supported in part by the National Key Research and Development Program of China under Grant 2020YFB1807201, in part by the National Natural Science Foundation of China under Grant 62031019, and in part by the National Science Foundation of China under Grant 62288101.

The authors would like to thank Zidong Wu, Yuhao Chen, Yashuai Cao, and Zhenchen Peng from Tsinghua University for their voluntary help during the field test experiments.

## Author contributions

J. Zhu conceived this idea and conducted the theoretical analysis. Z. Gu and Q. Ma conducted numerical simulations and fabricated the metasurface, J. Zhu and Z. Gu conducted the field test experiment. L. Dai and T. J. Cui directed and supervised the research. J. Zhu and Z. Gu processed the data and drafted the original manuscript. All authors participated in the data analysis and read the paper.

## Competing interests

The authors declare no competing interests.

## Additional information

**Correspondence** and requests for materials should be addressed to L. Dai or T. J. Cui.